\documentclass[twocolumn, preprintnumbers, showpacs, nofootinbib, aps, prl]{revtex4-1}
\usepackage{graphicx}

\newcommand{\bi}{\bibitem}
\newcommand{\be}{\begin{eqnarray}}
\newcommand{\ee}{\end{eqnarray}}

\begin{document}

\title{Can we constrain the maximum value for the spin parameter of the super-massive objects in galactic nuclei without knowing their actual nature?}

\author{Cosimo Bambi}
\email{Cosimo.Bambi@physik.uni-muenchen.de}
\affiliation{Arnold Sommerfeld Center for Theoretical Physics,
Ludwig-Maximilians-Universit\"at M\"unchen, 80333 Munich, Germany}
\affiliation{Institute for the Physics and Mathematics of the Universe, 
The University of Tokyo, Kashiwa, Chiba 277-8583, Japan}

\date{\today}


\begin{abstract}
In 4-dimensional General Relativity, black holes are described by 
the Kerr solution and are subject to the bound $|a_*| \le 1$, where 
$a_*$ is the black hole spin parameter. If current black hole candidates 
are not the black holes predicted in General Relativity, this bound 
does not hold and $a_*$ might exceed 1. In this letter, I relax the
Kerr black hole hypothesis and I find that the value of the spin parameter 
of the super-massive black hole candidates in galactic nuclei cannot 
be higher than about 1.2. A higher spin parameter would not be 
consistent with a radiative efficiency $\eta > 0.15$, as observed at least 
for the most luminous AGN. While a rigorous proof is lacking, I conjecture
that the bound $|a_*| \lesssim 1.2$ is independent of the exact nature 
of these objects.  
\end{abstract}


\maketitle


{\it Introduction ---}
Nowadays there is robust observational evidence of the existence
of $5 - 20$~$M_\odot$ dark bodies in X-ray binary systems and of 
$10^5 - 10^9$~$M_\odot$ dark bodies in galactic nuclei~\cite{ram}.
While the estimate of the masses of these objects is reliable,
as based on dynamical measurements, we do not know very much 
about their true nature.
The conjecture is that they are the black holes 
(BHs) predicted in General Relativity. The stellar-mass objects
in X-ray binary systems are too heavy to be neutron or quark stars
for any reasonable matter equation of state~\cite{bh1}. At least
some of the super-massive objects in galactic nuclei are too 
massive, compact, and old to be clusters of non-luminous bodies,
as the cluster lifetime due to evaporation and physical collision
would be shorter than the age of these systems~\cite{bh2}. 
However, constraints on the geometry of the space-time 
around these objects are weak~\cite{agn}. 
For the time being, we have to fully rely
on the validity of General Relativity, which is tested only in the 
weak field limit (Solar System and binary pulsars),
where $g_{tt} \approx - (1 + 2\phi)$ and $|\phi| \lesssim 10^{-6}$~\cite{will}.

In 4-dimensional General Relativity, BHs are described by the Kerr
solution and are completely specified by two parameters: the mass
$M$ and the spin angular momentum $J$~\cite{nohair}. 
A fundamental limit for a
BH in General Relativity is the Kerr bound $|a_*| \le 1$, where
$a_* = J/M^2$ is the spin parameter. This is just the condition
for the existence of the event horizon: for $|a_*| > 1$ the event
horizon disappears and the central singularity becomes naked,
violating the weak cosmic censorship conjecture~\cite{wccc}.
Despite the possibility of forming naked singularities from regular
initial data~\cite{pankaj}, the
existence of a Kerr naked singularity can be excluded at least
for the following reasons: it is apparently impossible to make a star 
collapse with $|a_*| > 1$~\cite{rez} or overspin an already existing
BH up to $|a_*| > 1$~\cite{eb1} and, even if created, a Kerr naked
singularity would be highly unstable~\cite{eb2}.

On the other hand, if the current BH candidates are not the BHs
predicted in General Relativity, the Kerr bound does not hold
and the maximum value for $a_*$ may be either larger or smaller
than 1, depending on the actual nature of these objects~\cite{super,super2}.
Generally speaking, bodies with spin parameter larger than 1
are not necessarily monsters: for non-compact objects, $a_*$
can easily exceed 1. For example, the Earth has $a_* \sim 10^3$. In
the case of compact objects, a high $a_*$ is more difficult and, for
instance, the maximum value of the spin parameter of a neutron 
star is thought to be about 0.6, because otherwise the object 
becomes unstable and
spins down by emitting gravitational radiation~\cite{ns}.
As shown in~\cite{evolution}, if the geometry around a 
compact object deviates from the Kerr metric, the accretion
process can naturally spin the object up to $|a_*| > 1$.

On the basis of these considerations, it is interesting to 
figure out if current observations can provide some constraint
on the maximum value of the spin of the BH candidates,
even if we do not know their nature.

{\it Non-Kerr compact objects ---}
At first approximation, a non-Kerr compact object can be
described by three parameters: in addition to the mass and
the spin angular momentum, we can introduce a ``deformation 
parameter'', say
$\epsilon$, which measures the deviations from the Kerr geometry. 
It is convenient that for $\epsilon = 0$ we recover exactly
the Kerr solution. In the literature there are a few proposals 
that can do the job~\cite{metric}. Here I use the metric recently
suggested in~\cite{ps}, because it has the advantage that
$a_*$ and $\epsilon$ can assume any value. In Boyer-Lindquist
coordinates, the metric reads
\be\label{gmn}
g_{tt} &=& - \left(1 - \frac{2 M r}{\Sigma}\right)
\left(1 + h\right) \, , \nonumber\\
g_{t\phi} &=& - \frac{2 a M r \sin^2\theta}{\Sigma} 
\left(1 + h\right) \, , \nonumber\\
g_{rr} &=& \frac{\Sigma \left(1 + h\right)}{\Delta 
+ a^2 h \sin^2 \theta} \, , \nonumber\\
g_{\theta\theta} &=& \Sigma \, , \nonumber\\
g_{\phi\phi} &=& \left(r^2 + a^2 + \frac{2 a^2 M r 
\sin^2\theta}{\Sigma}\right) \sin^2\theta + \nonumber\\
&& + \frac{a^2 (\Sigma + 2 M r) \sin^4\theta}{\Sigma} 
h \, , \nonumber\\
\ee
where $a = a_* M$ and
\be
\Sigma &=& r^2 + a^2 \cos^2\theta \, , \nonumber\\
\Delta &=& r^2 - 2 M r + a^2 \, \nonumber\\
h &=& \frac{\epsilon M^3 r}{\Sigma^2} \, .
\ee
The compact object is more 
prolate (oblate) than a Kerr BH for $\epsilon > 0$ ($\epsilon < 0$);
when $\epsilon = 0$, we recover the Kerr solution.

{\it Radiative efficiency ---}
The luminosity of a compact object due to the accretion
process is simply $L_{acc} = \eta \dot{M} $, where
$\eta$ is the radiative efficiency and $\dot{M}$ is the mass
accretion rate. The value
of $\eta$ depends on the details of the accretion process.
For instance, in the case of Bondi accretion onto a
Schwarzschild BH, the gas cannot radiate efficiently the
energy gained by falling into the BH gravitational potential
and $\eta \sim 10^{-4}$~\cite{book}. The highest value of the
radiative efficiency is reached when a BH 
is surrounded by a geometrically thin and optically thick
accretion disk. The gas's particles inspiral into the central
object by losing energy and angular momentum. When
they reach the inner edge of the disk, which can be supposed
to be at the innermost stable circular orbit (ISCO), 
they plunge into the BH.
If the gas does not emit additional radiation during the 
plunging and no radiation is emitted from the surface of the
compact object (as it is observed in the case of BH 
candidates~\cite{horizon}), the radiative efficiency is
\be\label{eta}
\eta = 1 - E_{\rm ISCO} \, ,
\ee
where $E_{\rm ISCO}$ is the specific energy of a particle
at the ISCO. In the Kerr background, $\eta = 0.057$ for a
non-rotating BH, $\eta = 0.32$ for a BH rotating at the
Thorne's limit (i.e. $a_* = 0.998$)~\cite{thorne}, and 
$\eta = 0.42$ for an extreme BH (i.e. $a_* = 1$).

For a generic axially symmetric and stationary background,
Eq.~(\ref{eta}) can be computed as follows.
One assumes that the disk is on the equatorial plane and that the
gas moves on nearly geodesic circular orbits. In cylindrical 
coordinates, the equations of the geodesic motion 
of a particle around the compact object are
\be
\dot{t} &=& \frac{Eg_{\phi\phi} + Lg_{t\phi}}{g_{t\phi}^2 - 
g_{tt}g_{\phi\phi}} \, , \\
\dot{\phi} &=& \frac{Eg_{t\phi} + Lg_{tt}}{g_{t\phi}^2 
- g_{tt}g_{\phi\phi}} \, , \\
g_{rr} \dot{r}^2 + g_{zz} \dot{z}^2 &=&
V_{\rm eff}(E,L,r,z) \, ,
\ee
where $E$ and $L$ are respectively the conserved specific
energy and the conserved specific $z$-component of the angular 
momentum, while $V_{\rm eff}$ is the effective potential 
\be
V_{\rm eff} = \frac{E^2g_{\phi\phi} + 2 E L g_{t\phi} + 
L^2g_{tt}}{g_{t\phi}^2 - g_{tt}g_{\phi\phi}} - 1 \, .
\ee
Circular orbits on the equatorial plane are located at the zeros
and the turning points of the effective potential: $\dot{r}=\dot{z}=0$
implies $V_{\rm eff} = 0$, and $\ddot{r}=\ddot{z}=0$ requires
$\partial_rV_{\rm eff}=\partial_zV_{\rm eff}=0$. From these conditions,
we can get $E$ and $L$:
\be
E &=& - \frac{g_{tt}+g_{t\phi}\Omega}{\sqrt{-g_{tt}-2g_{t\phi}
\Omega-g_{\phi\phi}\Omega^2}} \, , \\
L &=& \frac{g_{t\phi}+g_{\phi\phi}\Omega}{\sqrt{-g_{tt}-2g_{t\phi}
\Omega-g_{\phi\phi}\Omega^2}} \, ,
\ee
where
\be
\Omega = \frac{-\partial_r g_{t\phi} \pm 
\sqrt{(\partial_rg_{t\phi})^2 - 
(\partial_rg_{tt})(\partial_rg_{\phi\phi})}}{\partial_rg_{\phi\phi}}
\ee
is the orbital angular velocity and the sign $+$ ($-$) is for corotating
(counterrotating) orbits. The orbits are stable under small perturbation
if $\partial_r^2V_{\rm eff} \le 0$ and $\partial_z^2V_{\rm eff} \le 0$.
One can thus find numerically the ISCO radius and get the specific
energy $E_{\rm ISCO}$ and the maximum efficiency parameter
$\eta = 1 - E_{\rm ISCO}$ for any value of $a_*$ and $\epsilon$.

\begin{figure*}
\par
\begin{center}
\includegraphics[type=pdf,ext=.pdf,read=.pdf,width=8.5cm]{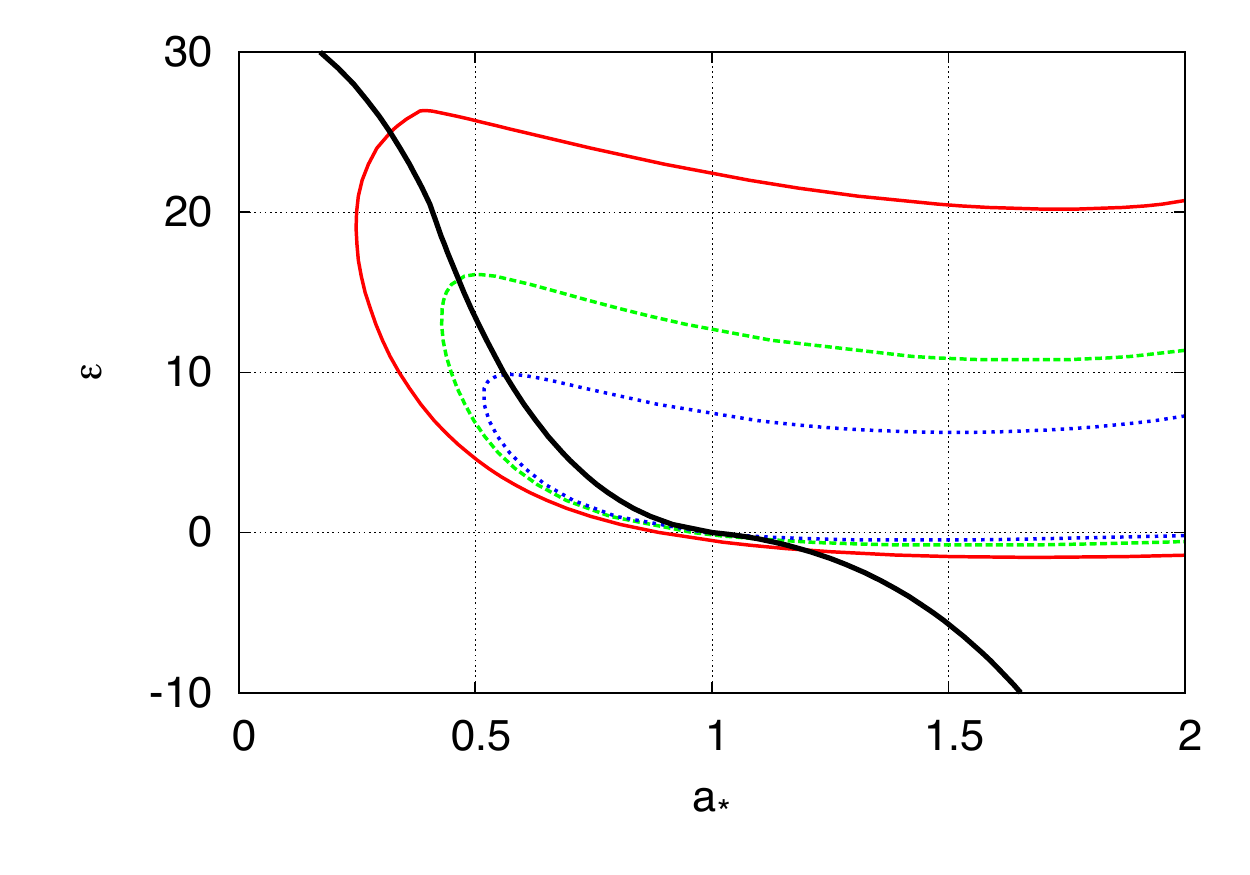}
\includegraphics[type=pdf,ext=.pdf,read=.pdf,width=8.5cm]{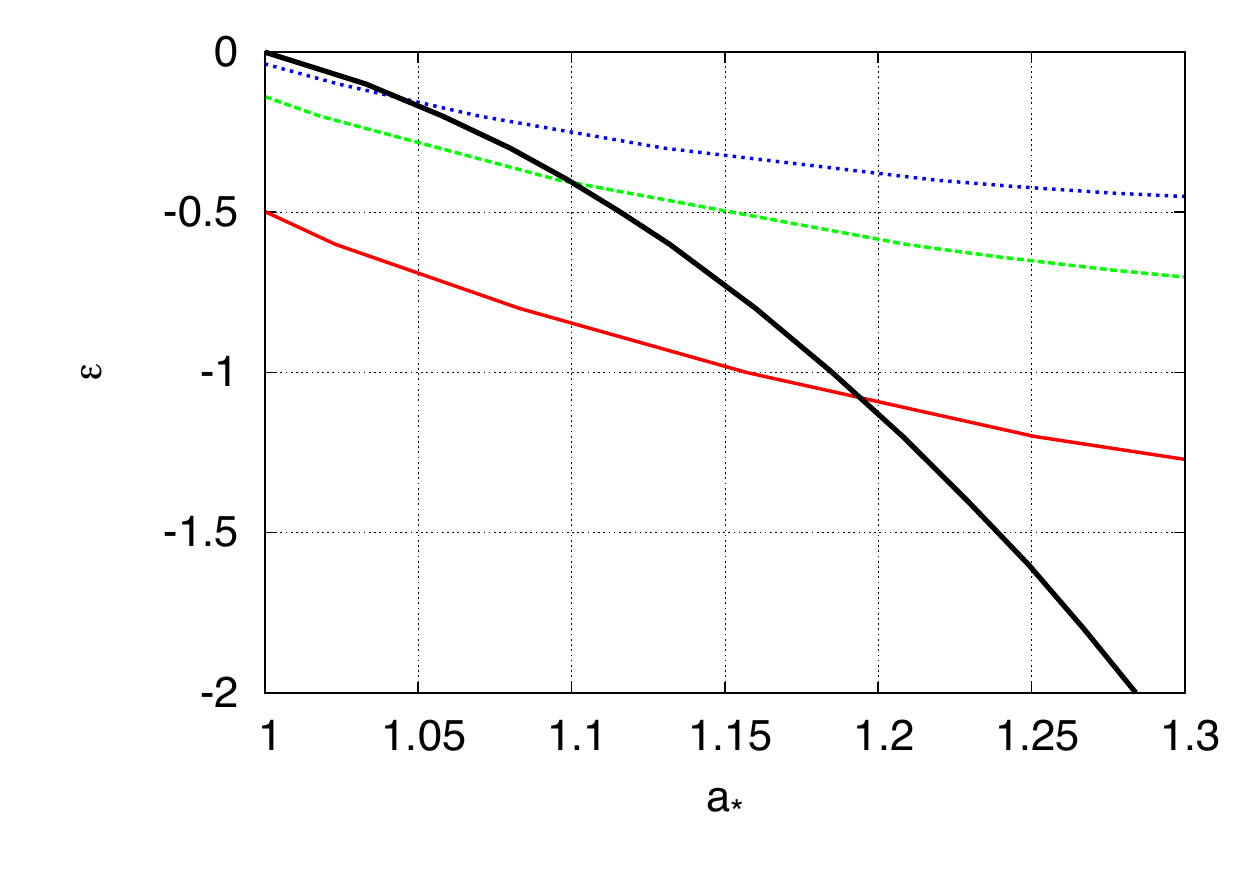}
\end{center}
\par
\vspace{-5mm} 
\caption{Compact objects with spin parameter $a_*$ and deformation
parameter $\epsilon$. The radiative efficiency is $\eta = 0.15$ 
(red solid curve), $\eta = 0.20$ (green dashed curve), and 
$\eta = 0.25$ (blue dotted curve). The black solid curve is the
equilibrium spin parameter $a_*^{eq}$ obtained from Eq.~(\ref{eq-a}).
The right panel is an enlargement of the parameter region $1.0 < a_* < 1.3$ 
and $-2.0 < \epsilon < 0.0$.}
\label{fig}
\end{figure*}

Fig.~\ref{fig} shows some contours of the radiative efficiency 
for an object with spin parameter $a_*$ and deformation 
parameter $\epsilon$ for the metric~(\ref{gmn}). The radiative 
efficiency is $\eta = 0.15$
(red solid curve), $\eta = 0.20$ (green dashed curve), and 
$\eta = 0.25$ (blue dotted curve).

{\it Evolution of the spin parameter ---}
The value of the spin parameter of a compact object is determined
by the competition of three physical processes: the event creating
the object, mergers, and gas accretion. For the super-massive objects
in galactic nuclei, independently of their nature, the initial value of the
spin parameter is completely irrelevant: their mass
has increased by several orders of magnitude from its original value,
and the spin parameter has evolved accordingly. On average, the 
capture of small bodies (minor merger) in randomly oriented orbits
spins any compact object down, since the magnitude of the orbital
angular momentum for corotating orbits is always smaller than the
one for counterrotating orbits~\cite{scott}. The case of coalescence of 
two compact objects with comparable mass (major merger) can be
rigorously computed only if we know the exact nature of these objects
and the theory of gravity, as the background is not fixed and the 
emission of gravitational waves may be important. In General Relativity,
the product of the merger of two neutron stars is a black hole with
$a_* \approx 0.78$, depending only weakly on the total mass and 
mass ratio of the system~\cite{kyoto}. In the case of random merger 
of two black holes, the most probable final product is a black hole with 
$a_* \approx 0.70$, while fast-rotating object with $a_* > 0.9$ should 
be rare~\cite{berti}.

Accretion from a disk can potentially be a very efficient way to spin
a compact object up~\cite{berti}. If the inner edge of the disk is at the 
ISCO radius, the
gas's particles plunge into the compact object with specific energy
$E_{\rm ISCO}$ and specific angular momentum $L_{\rm ISCO}$.
The mass $M$ and the spin angular momentum $J$ of the compact 
object change respectively by $\delta M = E_{\rm ISCO} \delta m$
and $\delta J = L_{\rm ISCO} \delta m$, where $\delta m$ is the
gas rest-mass. The evolution of the spin parameter is
governed by the following equation~\cite{bard}
\be\label{eq-a}
\frac{da_*}{d\ln M} = \frac{1}{M} 
\frac{L_{\rm ISCO}}{E_{\rm ISCO}} - 2 a_* \, ,
\ee
neglecting the small effect of the radiation emitted by the disk and
captured by the object. If accretion proceeds via short episodes
(chaotic accretion)~\cite{king}, the net effect is not different from
minor mergers in randomly oriented orbits and the compact object is spun down.
On the contrary, prolonged disk accretion is a very efficient mechanism
to spin the compact object up, till an equilibrium spin parameter
$a_*^{eq}$, which is reached when the right-hand side of Eq.~(\ref{eq-a})
becomes zero. For instance, an initially non-rotating Kerr BH reaches
the equilibrium $a_*^{eq} = 1$ after having increased its mass by
a factor $\sqrt{6} \approx 2.4$~\cite{bard}.

We can thus say that the most optimistic scenario to produce
fast-rotating super-massive objects at the center of galaxies is via
prolonged disk accretion and that the maximum value for the spin
parameter of these objects cannot exceed $a_*^{eq}$. The numerical
value of $a_*^{eq}$ depends on the 
metric of the space-time. The black solid curve in Fig.~\ref{fig}
shows the equilibrium spin parameter for the metric~(\ref{gmn}).
Objects on the left of the
black solid curve have $a_* < a_*^{eq}$ and the accretion 
process spins them up; objects on the right have $a_* > a_*^{eq}$
and the accretion process spins them down.
As already noted in Ref.~\cite{evolution}, objects more oblate 
than a BH (for the metric~(\ref{gmn}) when $\epsilon < 0$) have 
$a_*^{eq} > 1$.

{\it Observational constraints ---}
In general, it is not easy to get an estimate of $\eta$, as the
measurement of the mass accretion rate $\dot{M}$ is typically
quite problematic. The mean radiative efficiency of AGN can
be inferred from the Soltan's argument~\cite{soltan}, which relates
the mass density of the super-massive BH candidates in the
contemporary Universe with the energy density of the radiation
produced in the whole history of the Universe by the accretion
process onto these objects. There are several sources of uncertainty
in the final result, but a mean radiative efficiency $\eta > 0.15$
seems to be a conservative lower limit~\cite{elvis}.
The authors of Ref.~\cite{ho} find a mean radiative efficiency
$\eta \approx 0.30 - 0.35$ without some important assumptions
necessary in the original version of the Soltan's argument.
In Ref.~\cite{davis}, the authors show how to estimate $\eta$
for individual AGN and find that the more massive objects
have typically higher $\eta$, up to $\sim 0.3 - 0.4$.

Here, it is not important the mean radiative efficiency
of these objects. It is sufficient to say that at least some of them
must have $\eta > 0.15$. In other words, the space-time around
the super-massive BH candidates allows for a specific energy
at the ISCO radius smaller than 0.85. This fact is non-trivial,
as it says that the ISCO radius can be quite close to the object
(= gravity cannot be too strong). On the other hand, very high spin
parameters could be possible only in stronger gravitational
fields, in which the ISCO radius is larger and $L_{\rm ISCO}/E_{\rm ISCO}$
is larger too. So, if we use the metric~(\ref{gmn}) to describe the
geometry of the space-time around the super-massive BH 
candidates in galactic nuclei, we find that the spin parameter
of these objects cannot exceed 1.19, see Fig.~\ref{fig}. The 
maximum value for $|a_*|$ becomes 1.10 if we require $\eta > 0.20$,
and 1.04 for $\eta > 0.25$.

\begin{figure}
\par
\begin{center}
\includegraphics[type=pdf,ext=.pdf,read=.pdf,width=8.5cm]{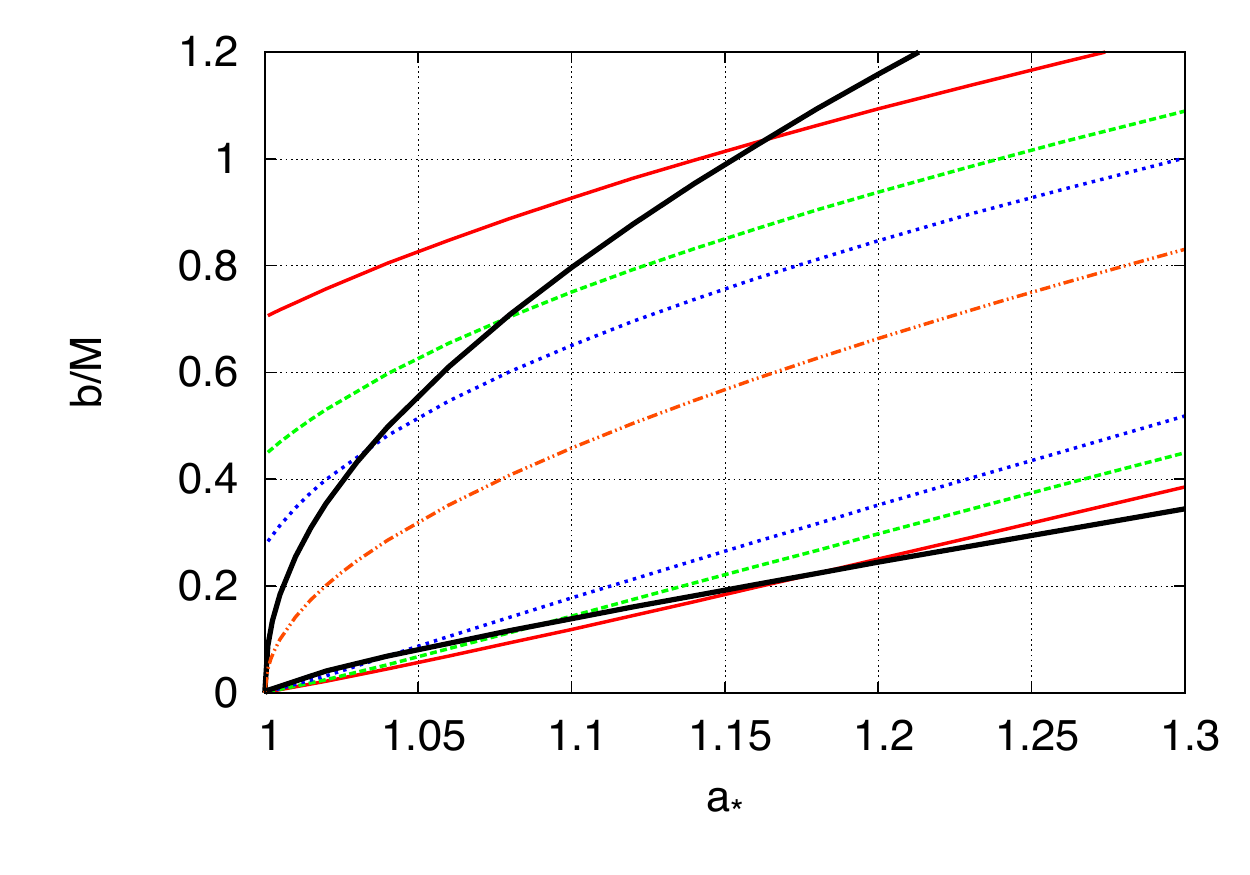}
\end{center}
\par
\vspace{-5mm} 
\caption{Compact objects with spin parameter $a_*$ and deformation
parameter $b$ described by the Manko-Mielke-Sanabria Gomez (MMS) 
solution. The radiative efficiency is $\eta = 0.15$ (red solid curve), 
$\eta = 0.20$ (green dashed curve), and $\eta = 0.25$ (blue dotted curve). 
The black solid curve is the equilibrium spin parameter $a_*^{eq}$ 
obtained from Eq.~(\ref{eq-a}). For $b = M \sqrt{a_*^2 - 1}$ (orange 
dashed-dotted curve), we recover the Kerr solution, which is in the
region $a_* > a_*^{eq}$.}
\label{fig2}
\end{figure}

{\it Comments ---}
To show that the result $|a_*| \lesssim 1.2$ seems to be robust, 
it is necessary
to address at least two points, concerning respectively its dependence on 
the choice of the metric~(\ref{gmn}) and the validity of Eqs.~(\ref{eta})
and (\ref{eq-a}).

The bound $|a_*| \lesssim 1.2$ seems to depend only marginally
on the choice of the metric. For instance, we get quite similar 
constraints if we consider the Manko-Mielke-Sanabria Gomez (MMS) 
metric, which is an exact solution of the Einstein's vacuum equation
(the metric~(\ref{gmn}) is not) and does not describe a BH (while the 
metric~(\ref{gmn}) does); see the second paper in~\cite{evolution}.
In addition to the mass and the spin parameter, the MMS solution
has a deformation parameter $b$. When $|a_*| \ge 1$, the Kerr metric
is recovered for $b = \pm M \sqrt{a_*^2 - 1}$; around $b = M 
\sqrt{a_*^2 - 1}$ there are objects more oblate than Kerr BHs, 
around $b = - M \sqrt{a_*^2 - 1}$ the objects are more prolate than
Kerr BHs. The constraints on the maximum value for the spin parameter
are shown in Fig.~\ref{fig2} -- here I show only the parameter space
$b > 0$ because for more prolate objects we find lower values. The
bounds turn out to be 1.18 if we require $\eta > 0.15$, 1.09 for 
$\eta > 0.20$, and 1.04 for $\eta > 0.25$. Despite the different nature 
of the two metrics, it is remarkable that we get very similar constraints.
The point is that the constraint on the maximum value for the spin 
parameter is not very sensitive to the exact space-time geometry, but
it depends on how much the compact object is more or less oblate.

The result relies also on the validity of Eqs.~(\ref{eta}) and (\ref{eq-a}).
As discussed in Ref.~\cite{prep}, in backgrounds deviating from
the Kerr geometry, the gas may not plunge from the ISCO into the
central object; if this is the case, the gas must 
form a thick disk inside the ISCO radius
and lose additional energy and angular momentum. That increases
the radiative efficiency at most by a few percent with respect to the
value calculated from Eq.~(\ref{eta}). It also slightly decreases the
equilibrium parameter $a_*^{eq}$, as the gas plunges from a radius
inside the ISCO. However, for the metric~(\ref{gmn}) and 
$\epsilon < 0$ such a possibility never happens: accretion 
proceeds as in the Kerr space-time and the result $|a_*| \lesssim 1.2$
is not affected.

Lastly, let us consider the possibility that the initial value of the spin
parameter of the object is $a_*^{in} > a_*^{eq}$. In this case, the accretion 
process would spin the object down, approaching $a_*^{eq}$ from 
the right of the black solid line in Fig.~\ref{fig}, but the bound $|a_*| 
\lesssim 1.2$ can still be applied. The initial value of 
the spin parameter of the super-massive objects in galactic nuclei is 
presumably negligible: their mass has increased by several orders 
of magnitude from its original value and $a_*$ has evolved according 
to Eq.~(\ref{eq-a}). If $a_*^{in}$ were of order 1, $a_*^{eq}$ was reached
soon, after the object increased its mass by a factor of order 1. 
The possibility that this gravity theory can make a star collapse with 
$|a_*| \gg 1$ and that the super-massive black hole candidates have 
still a spin parameter significantly larger than $a_*^{eq}$ seems to be very 
unlikely, at least for two reasons. The accretion process onto an object 
with $|a_*| \gg 1$ is strongly suppressed and the object could have 
not become super-massive~\cite{super2}. This behavior does not 
depend on the exact metric of the space-time, because the effect of 
the spin would be important relatively far from the object, where
deviations from the Kerr geometry are more suppressed.
The second reason is that compact 
objects with spin parameter $|a_*| \gg 1$ are usually unstable. For instance,
Ref.~\cite{eb2} shows that the Kerr metric with $|a_*| > 1$ is unstable because 
of the existence of stable photon orbits with negative energy and that this 
conclusion does not depend on the exact gravity theory.

{\it Conclusions ---}
In 4-dimensional General Relativity, BHs are subject to the bound
$|a_*| \le 1$, where $a_* = J/M^2$ is the spin parameter. If the
current BH candidates are not the BH predicted in General Relativity,
this bound does not hold and $a_*$ might exceed 1. In this letter,
I have relaxed the common assumption that the super-massive
objects at the center of every normal galaxy are Kerr BHs and
I have found that current observations can constrain the maximum 
value of the spin parameter of these object at the level of 
$|a_*| \lesssim 1.2$. While I cannot provided a rigorous proof, 
my conjecture is that this bound holds whatever the nature of these 
objects is. The origin of this bound can be heuristically understood 
as follows. A higher spin parameter would require a larger
ISCO radius: both $L_{\rm ISCO}$ and $E_{\rm ISCO}$ decrease 
as the ISCO radius decreases, but $L_{\rm ISCO}$ decreases 
faster. However, a larger ISCO radius implies a lower maximum
radiative efficiency $\eta_{max} = 1 - E_{\rm ISCO}$, while we
know that at least some of the super-massive objects in galactic
nuclei must have $\eta > 0.15$. From the latter, we get
$|a_*| \lesssim 1.2$.


\begin{acknowledgments}
This work was supported by World Premier International Research 
Center Initiative (WPI Initiative), MEXT, Japan, by the JSPS 
Grant-in-Aid for Young Scientists (B) No. 22740147, and
by Humboldt Foundation. 
\end{acknowledgments}


\end{document}